**Automated Classification of Volcanic Earthquakes Using Transformer Encoders: Insights into Data Quality and Model Interpretability**

Y. Suzuki[1], Y. Yukutake[2], T. Ohminato[2], M. Yamasaki[1] and Ahyi Kim[1*]

[1]Yokohama City University

[2]Earthquake Research Institute, The University of Tokyo

Corresponding author: Ahyi Kim (ahyik@yokohama-cu.ac.jp)

22 Seto Kanazawa-ku Yokohama city Kanagawa prefecture Japan, 236-0027

**Abstract**

Precisely classifying earthquake types is crucial for elucidating the relationship between volcanic earthquakes and volcanic activity. However, traditional methods rely on subjective human judgment, which requires considerable time and effort. To address this issue, we developed a deep learning model using a transformer encoder for a more objective and efficient classification. Tested on Mount Asama's diverse seismic activity, our model achieved high F1 scores (0.930 for volcano tectonic, 0.931 for low-frequency earthquakes, and 0.980 for noise), superior to a conventional CNN-based method. To enhance interpretability, attention weight visualizations were analyzed, revealing that the model focuses on key waveform features similarly to human experts. However, inconsistencies in training data, such as ambiguously labeled B-type events with S-waves, were found to influence classification accuracy and attention weight distributions. Experiments addressing data selection and augmentation demonstrated the importance of balancing data quality and diversity. In addition, stations within 3 km of the crater played an important role in improving model performance and interpretability. These findings highlight the potential of Transformer-based models for automated volcanic earthquake classification, particularly in improving efficiency and interpretability. By addressing challenges such as data imbalance and subjective labeling, our approach provides a robust framework for understanding seismic activity at Mount Asama. Moreover, this framework offers opportunities for transfer learning to other volcanic regions, paving the way for enhanced volcanic hazard assessments and disaster mitigation strategies.

**Introduction**

Active earthquake swarms are frequently associated with volcanic activity in volcanic regions. Monitoring these earthquakes is crucial for assessing current volcanic activity and providing insights into the physical processes of volcanic fluids, such as magma and hydrothermal fluids (Chouet *et al.*, 1988; McNutt, 1996; Nishimura and Iguchi, 2011). Minakami *et al.* (1970) proposed that volcanic earthquakes could be classified into four types based on waveform characteristics, source location, and their relationship with surface activity: A-type (tectonic earthquakes), B-type earthquakes (commonly referred to as low-frequency earthquakes, although shallow earthquakes at Mount Asama can be classified as B-type regardless of their waveform characteristics, implying some B-type events do not strictly exhibit low-frequency features), volcanic tremors, and explosion earthquakes. Although the frequency components of these four types of earthquakes vary by volcano and may be further subdivided or include other types, a typical pattern before eruptions begin is an increase in A-type earthquakes, followed by an increase in B-type earthquakes and volcanic tremors (Oikawa *et al.*, 2006; Iguchi, 2013). We note A- and B-type classifications commonly used in Japan correspond to Volcano-Tectonic (VT) events and Long-Period (LP) events, respectively, as is more common internationally (Chouet, 1996, 2003). These seismic events often indicate changes in subsurface magmatic and hydrothermal activity, making them crucial for monitoring volcanic hazards. However, the exact relationship between these earthquakes and their subsequent eruptions remains unclear, particularly due to inconsistencies in classification methods. Developing precise classification criteria for A-type and B-type earthquakes is essential not only for improved understanding of these processes, but also for enhancing eruption forecasting and developing more effective disaster mitigation strategies.

Traditional methods classify volcanic earthquakes based on visual observations of the seismic wave amplitude, dominant frequency, duration, and related surface activity evaluated by experts, which are time-consuming and expensive. Additionally, the classification criteria may need to be standardized across observers and institutions.

Permana et al. (2022) aimed to automate classification specifically for volcanic tremors and B-type earthquakes based on seismic correlation analysis, distinguishing them from continuous seismic recordings. Consequently, they analyzed continuous data from six seismic stations at Sakurajima Volcano, Japan, from April to September 2017, correctly classifying approximately 80% of the volcanic earthquakes and 75% of the B-type earthquakes reported in the Japan Meteorological Agency catalog. However, this method relies on the correlation of seismic waveforms, and the issue of reduced classification accuracy persists for signals impacted by noise or lacking sufficient coherence.

Several machine learning models have also been proposed for the automatic classification of volcanic earthquakes and tremors. Unsupervised learning methods include Hidden Markov Models (Bebbington, 2007), Self-Organizing Maps (SOM) (Langer et al., 2009; Köhler et al., 2010; Langer et al., 2011), Gaussian Mixture Models (Hammer et al., 2012), and Principal Component Analysis (PCA) (Unglert et al., 2016). Ida et al. (2022) developed a method combining k-means clustering and Dynamic Time Warping (DTW) to classify volcanic earthquake waveforms, applying it to data from the day when magma ascent was observed beneath the active crater of Sakurajima Volcano, demonstrating its validity. However, unsupervised learning faces challenges in interpretations and the need for manual labeling by human experts in the final classification.

In contrast, supervised learning models include the work of Malfante et al. (2018), who used Support Vector Machines (SVMs) to classify six types of seismic events with 92% accuracy. Titos et al. (2018) employed Convolutional Neural Networks (CNNs) to classify volcanic earthquakes, demonstrating that these methods can be applied even in areas with relatively few observational data through transfer learning. Due to the significant influence of source mechanisms on spectral characteristics, some studies have used time-frequency representations with 2-D CNNs (e.g., Canário et al., 2020; Lara et al., 2021). Nakano et al. (2022) showed that there is no significant difference in performance between the two, because even in a 1-D CNNs with only a temporal domain, frequency components are accounted for when extracting features in the convolution layer. While deep learning has significantly improved the performance of earthquake type classification and reduced human effort and time costs, the lack of transparency in the classification rationale remains a challenge.

In this study, we propose a method using a Transformer encoder to overcome these challenges. The Transformer can learn features directly from raw data, greatly simplifying the preprocessing and feature extraction process; its flexibility allows it to adapt to different volcanic and seismic environments. Additionally, by visualizing the Attention Weights (AW), we aimed to render a portion of the classification process interpretable, enabling us to examine whether the model follows similar criteria to human experts or if it uses unique classification criteria. The use of Transformer architectures in seismology has been explored previously by Mousavi et al. (2020), who developed EQTransformer for simultaneous earthquake detection and phase picking. While their model focuses on tectonic earthquakes, our study employs a transformer encoder specifically for classifying volcanic earthquakes, addressing the unique characteristics of A- and B-type events and enhancing interpretability for volcanic monitoring purposes.

**Methods**

Model Architecture Overview

We developed a neural network for classifying seismic events that combines a one-dimensional convolutional front-end with a Transformer encoder backend (Figure 1). The input to the model is a three-component seismic waveform segment of length 3,000 samples (30 s at 100 Hz). First, a one-dimensional convolution layer with a kernel size of 50, stride of 10, input channels = 3 (for the three components EW, NS, UD), and output channels = 150 processes the waveform. This convolution segments the waveform into overlapping 50-sample frames and outputs 150 features per frame. With the given stride, the convolution produces a sequence of 296 feature vectors of length 150 each, preserving temporal order while reducing sequence length for efficiency. We apply batch normalization and a ReLU activation to the convolutional outputs. By using a larger kernel to capture longer temporal patterns and a larger stride (instead of a pooling layer) to achieve dimension reduction, we retain important temporal information while efficiently reducing sequence length, similar to the approach in Gulati et al. (2020).

Next, we insert a special classification token (CLS) at the beginning of the sequence of convolutional features, immediately following positional embedding (Figure 1), as a learnable summary representation for classification. The sequence length thereby becomes 297 (296 feature vectors plus one CLS token). This sequence (CLS + features) is fed into three stacked Transformer encoder layers. Finally, the Transformer output from the CLS token embedding is passed through a fully connected layer (linear layer) followed by a softmax function to obtain class probabilities for A-type, B-type, or noise event classes. Without the CLS token, the model would need to aggregate sequence information through alternative methods such as average or maximum pooling over all feature vectors. However, these pooling methods might discard critical contextual information needed for accurate classification. The CLS token explicitly provides a dedicated, learnable summary representation, allowing the model to effectively capture and integrate global sequence information via the self-attention mechanism. Prior to model training, we prepared seismic waveform data from Mount Asama, consisting of A-type and B-type earthquake waveforms, as well as noise segments (see Figure 2 for waveform examples and event distribution). Training was conducted for 60 epochs using a mini-batch size of 200 and a multiclass cross-entropy loss function. We utilized the AdamW optimizer (Loshchilov & Hutter, 2017), which incorporates weight decay into Adam (Kingma & Ba, 2014), with a learning rate of 0.001 and weight decay of 0.01. These hyperparameters were selected through parameter tuning to balance model performance and computational efficiency. The mini-batch size was set to the maximum feasible value that could fit within the GPU memory, ensuring stable and efficient training. The learning curves shown in Figure 3 represent the entire 60 epochs of training, demonstrating consistent convergence without overfitting. Note that the CLS token is a single vector with the same dimensionality (150) as each of the convolutional feature vectors. This single CLS vector is prepended once at the very beginning of the entire sequence (not separately to each of the 150 channels), resulting in a sequence of dimension 297×150.

Transformer Encoder and Self-Attention

Our model uses only the encoder component of the Transformer architecture (Vaswani et al., 2017), as our task involves encoding input sequences into representative vectors for classification. A key feature of the Transformer encoder is the self-attention mechanism, which weights the relevance of every input position relative to others. Each token (vector) $x_i$ (dimention $d_{model} = 150$) is projected into three vectors: Query ($q_i$), key ($k_i$), and value ($v_i$):

$$q_i = x_i W^Q, \quad k_i = x_i W^K, \quad v_i = x_i W^V \tag{1}$$

where $W^Q, W^K, W^V$ are parameter matrices each of dimension (150 × 15), with 150 corresponding to $d_{model}$ and the column size of 15 determined by dividing $d_{model}$ by the number of attention heads (10 heads), thus 150/10=15. The vectors, $q_i, k_i, v_i$ form matrices $Q$, $K$, $V$ respectively, each with dimensions (297 × 15):

$$Q, K, V \in \mathbb{R}^{L \times d_{model}} \tag{2}$$

where $L$=297 is the sequence length (296 convolutional feature vectors plus 1 CLS token), and $d_{model} = 150$. The attention weights computed from the dot product of Q and K thus form a square matrix of dimensions ($L \times L$) = (297×297).

Intuitively, $q_i$ represents a query (what information the token seeks), $k_i$ a key (the information held by each token), and $v_i$ the value (the actual content at each token). These projected vectors ($q_i, k_i, v_i$) and subsequent attention calculations are performed within the "Multi-Head Attention" block of the Transformer encoder, as illustrated in Figure 1. We set 10 attention heads ($h = 10$), making each head's dimension $d_k = 15$, since $h \times d_k = 150$. Multi-head attention enables capturing diverse aspects of information simultaneously, as each head learns to focus on different patterns or features within the waveform data.

The attention score $e_{ij}$ between tokens $i$ and $j$ is calculated by the scaled dot product of Query and Key vectors:

$$e_{ij} = \frac{q_i \cdot k_j}{\sqrt{d_k}} \quad (3)$$

These scores are normalized by the softmax function to obtain attention weights $a_{ij}$:

$$a_{ij} = \frac{\exp(e_{ij})}{\sum_{m=1}^{L} \exp(e_{im})} \quad (4)$$

where $L$ is the sequence length, which in this study is 297 (296 feature vectors from convolution plus 1 CLS token).

The scaling by $\sqrt{d_k}$ in the attention calculation prevents excessively large dot-product values ($q_i \cdot k_i$) as $d_k$ increases. This ensures that the softmax function does not produce overly peaked distributions, which would otherwise result in saturated gradients and unstable training (Vaswani et al., 2017).

Finally, the attention output $z_i$ for token $i$ is the weighted sum of Value vectors:

$$z_i = \sum_{j=1}^{L} a_{ij} v_{ij} \quad (5)$$

In matrix form, this is represented concisely as:

$$Attention(Q, K, V) = softmax\left(\frac{QK^T}{\sqrt{d_k}}\right)V \qquad (6)$$

Multi-Head Attention:

Our model employs multi-head attention with 10 heads to simultaneously capture multiple aspects of information. Each head performs the attention mechanism independently with its own projection matrices $W_m^Q, W_m^K, W_m^V$, each sized $150 \times 15$. Each head's output (size $L \times 15$) is concatenated into an $L \times 150$ matrix and linearly transformed by $W^O (150 \times 150)$:

$$MultiHead(Q, K, V) = Concat(head_1, \cdots, head_{10})W^O \qquad (7)$$

Each Transformer encoder layer also includes a position-wise feed-forward network (FFN), consisting of two linear layers with a ReLU activation:

$$FFN(x) = ReLU(xW_1 + b_1)W_2 + b_2 \qquad (8)$$

where $W_1$ is $150 \times 2048$, $W_2$ is $2048 \times 150$, $b_1$ is $2048 \times 1$, and $b_2$ is $150 \times 1$. Both the attention and FFN sub-layers utilize residual connections and layer normalization:

$$Output = LayerNorm\ (x + SubLayer(x)) \tag{9}$$

where SubLayer is either Multi-head Attention or FFN.

Positional Information and Relative Positional Embedding

Since the Transformer encoder does not inherently capture token order, we incorporate positional information via relative positional embeddings (Huang et al., 2018). We adopted this relative positional embedding because, for seismic waveform classification, capturing the relative timing between waveform features (such as the delay between observed P-waves and S-waves) is critical, whereas absolute timestamps within the waveform window are less informative. A trainable embedding vector $r_k$ represents each possible relative displacement $k$. This relative positional embedding vector $r_k$ has dimension $d_k = 15$, matching the dimension used for queries and keys within each attention head. A relative positional bias term $S_{i,j}^{rel}$ is computed as the dot product between the query vector qi at token position I and the relative positional embedding vector $r_{j-i}$, corresponding to the displacement $j - i$:

$$S_{i,j}^{rel} = q_i \cdot r_{j-i} \tag{10}$$

This bias term $S_{i,j}^{rel}$ is then added to attention scores based on token positions i and j. To efficiently construct $S^{rel}$, we use the Skew algorithm introduced by Huang et al. (2018). The Skew algorithm first pads the matrix obtained by multiplying queries by relative positional embeddings with an additional dummy column. Next, this padded matrix is flattened into a vector. A dummy row of length $N-1$ is then appended. Finally, the resulting vector is reshaped into a matrix of size $(N+1, 2N-1)$, and each subsequent row is shifted (skewed) by one position relative to the previous row. After removing excess elements introduced during padding, the resulting matrix $S^{rel}$ provides efficient computation of relative positional biases without explicitly calculating each pairwise positional displacement. The modified attention mechanism with relative position embedding is thus:

$$Attention_{rel}(Q, K, V) = softmax\left(\frac{QK^T + S^{rel}}{\sqrt{d_k}}\right)V \quad (11)$$

In this study, positional information is captured exclusively through these learned relative positional embeddings, without absolute positional encoding.

Classification Token and Output Layer

To aggregate sequence information into a single vector, we prepend a classification token (CLS). Unlike pooling, the CLS token is learnable and dynamically aggregates relevant information. Initially, the CLS embedding is the average of the convolutional outputs, then refined by Transformer layers. After the final encoder layer, the CLS token output vector (150-dimensional) is passed through a fully connected layer (150 × 3), producing logit scores for the three classes (A-type, B-type, noise). A softmax function converts these logits into class probabilities. Training uses cross-entropy loss, and at inference time, the class with the highest probability is selected. No reshaping or pooling of features is needed before Transformer input.

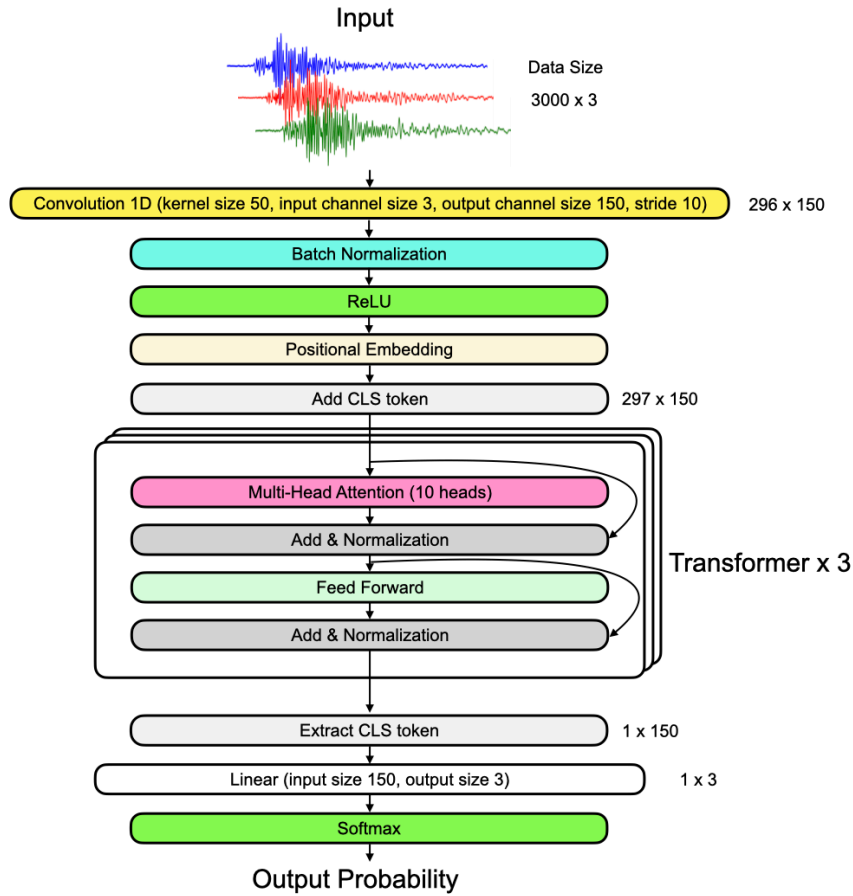

**Figure 1.** Network architecture. Three-component seismic waveforms are passed through a one-dimensional convolution layer (kernel size: 50; input channel size: 3; output channel size: 150; stride: 10), followed by batch normalization, ReLU activation, and positional embedding. A classification token (CLS) is added to the input sequence and processed through three stacked Transformer encoder layers, each consisting of 10 multi-head attention mechanisms, feedforward

sublayers, and residual connections. The CLS token output is extracted and passed through a linear transformation to classify A- and B-type events, as well as noise.

**Data**

To validate the classification model, we selected Mount Asama, an active andesitic volcano in central Japan, as our test site (Figure 2a). This volcano is known for frequent occurrences of both A-type (tectonic) and B-type (low-frequency) earthquakes, making it an ideal site to evaluate the model's ability to distinguish between these types of seismic events. Historical studies (Oikawa et al., 2006; Takeo et al., 2022) show that Mount Asama exhibits a typical pre-eruption seismic pattern, with an increase in A-type earthquakes followed by B-type events. The frequent occurrences of both earthquake types allow us to test the accuracy of our model in classifying seismic signals. Seismic observations have been conducted on Mount Asama for more than 100 years. Since the establishment of a modern observation network around the volcano in 2003, a network consisting of 30 seismometers, including 19 with broadband sensors, has been installed by institutions such as the Earthquake Research Institute (ERI), Japan Meteorological Agency (JMA), and National Research Institute for Earth Science and Disaster Resilience (NIED) (Figure 2a) up to 2017.

Volcanic earthquakes on Mount Asama have been classified based on their source locations and waveform characteristics (Minakami et al., 1970). Recent observations and research have identified several types of volcanic earthquakes on Mount Asama. B-type earthquakes, typically located directly beneath the crater, are characterized by lower frequency components and unclear P- and S-wave onsets. A-type earthquakes, situated slightly westward and at greater depths than the B-type distribution area, exhibit clearer P- and S-wave onsets and higher frequency components (Figure 2). Additionally, F-type earthquakes occur on the flanks of the volcano away from the crater while N-type earthquakes occur within the B-type source area, but are distinguished by decaying oscillatory waveforms (Oikawa et al., 2006). Currently, these events are further categorized into 12 specific types. However, A- and B-type earthquakes are of particular interest because they are commonly observed at many volcanoes and often follow a pre-eruption pattern in which A-type activity increases first, followed by an increase in B-type events (Oikawa et al., 2006). Thus, this study focused on classifying A-type and B-type earthquakes to establish a robust classification model that can serve as a basis for future generalization studies to other volcanoes while also contributing to a broader understanding of Mount Asama's seismic activity. Since our model was trained and tested exclusively on three waveform categories (A-type, B-type, and noise), waveforms belonging to other seismic event types identified at Mt. Asama (e.g., the other ten types) were not included in this study, partly due to their limited sample sizes. Thus, potential misclassification for these event types has not yet been assessed. Incorporating additional event types into future training datasets would be an important next step toward developing a more generalizable classification model.

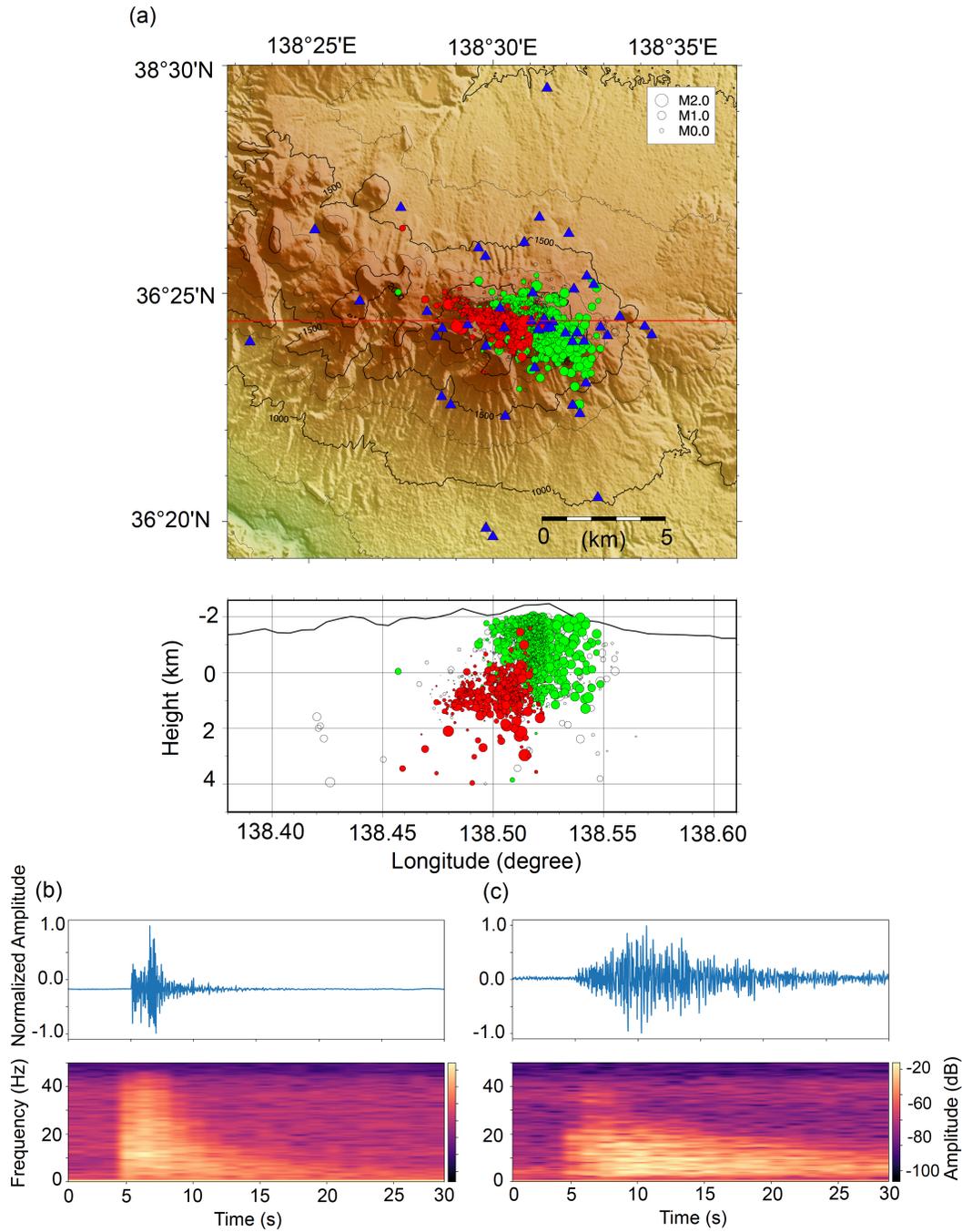

**Figure 2.** Distribution of earthquakes and stations and examples of input data. Earthquakes detected between January 2003 and October 2022, only with location determination errors less than 100 m in horizontal and depth directions (based on the earthquake catalog from the Earthquake Research Institute, The University of Tokyo), were plotted. (a) Distribution of earthquakes (circles) and stations (blue triangles) at Mount Asama. Red circles represent A-type events, green circles represent B-type labeled events, and open circles represent earthquakes that do not belong to either type. The top image is a map view, and the bottom image is a cross-sectional view along the red line in the map view. In the cross-section, "Height" represents elevation relative to sea level, with positive values indicating depth below the surface. The "0" on the vertical axis indicates the surface level. (b) An example of an A-type earthquake. The top panel shows the seismic waveform, and the bottom image shows the spectrogram. (c) Same as (b), but representing an example of a B-type earthquake.

Figure 2b and c respectively show waveforms and spectrograms of A- and B-type earthquakes.. A-type earthquakes have distinct P- and S-waves, whereas B-type earthquakes often have unclear P-wave onsets and indistinguishable S-waves. B-type earthquakes tend to have longer durations and lower dominant frequencies than A-type earthquakes. Therefore, this study aimed to classify A- and B-type earthquakes and differentiate them from noise. We used a seismic catalog between January 2003 and October 2022 developed by the ERI Asama Volcano Observatory to construct the classification model. The earthquake types in this catalog were manually classified based on visual waveform inspection. Figure 2a shows only those with horizontal and depth source errors within 100 m for 900 A- and 3,191 B-type earthquakes. To ensure comprehensive model training, all detected earthquakes, including those with relatively large location errors, were used. This comprised 1,027 A-type earthquakes (8,318 waveforms) and 4,090 B-type earthquakes (37,196 waveforms), with P- and S-wave arrival times manually selected. As shown in Figure 2, the volcanic earthquakes at Mount Asama, especially B-type earthquakes, are concentrated in shallow regions in the vicinity of the crater. Therefore, most seismic waveforms used in this study were observed at approximately 20 stations located within 3 km from the crater. However, the number of stations historically varied due to differences in operational duration, relocations, and data availability. Such variability, combined with issues of waveform quality and operational gaps at these stations, explains the discrepancy between the actual number of waveforms used and the theoretical maximum. The onsets of the P- and S-wave arrival times for these earthquakes were manually selected. Each dataset consisted of three channels (two horizontal and one vertical). We decimated the waveforms of stations recorded at 200 Hz such that the sampling frequency of all waveforms was 100 Hz. Each waveform was initially extracted from a segment of 9,001 points. For A- and B-type earthquakes, the starting point was randomly selected from 150 to 1,000 points before the P-wave arrival time, and 3000 points were extracted for use. Noise data were obtained from time interval with no earthquake signals. All waveform data were standardized before model training and testing by subtracting the mean and dividing by the standard deviation.

**Results**

*Model performance*

To construct the dataset, we included 112,852 noise waveforms derived from segments prior to the P-wave arrival in seismic records. This number exceeds the total waveforms for A- and B-type earthquakes because it also includes waveforms from events that do not belong to either class (A- or B-type). Importantly, the noise dataset does not include any waveforms overlapping with A-type, B-type, or any other categorized seismic events. For model evaluation, a fixed test set consisting of 1,663 waveforms for each class (A-type, B-type, and noise) was used across all experiments. The remaining waveforms were used for training. Detailed information on the dataset composition is provided in Table S1. The training curves (Figure 3) indicate that learning progressed without overfitting.

The model's performance was compared to that of a CNN-based model constructed following the architecture used by Nakano et al. (2022), consisting of four CNN layers and two fully connected layers (Figure S1). This CNN model was trained entirely using our dataset (Table 1).

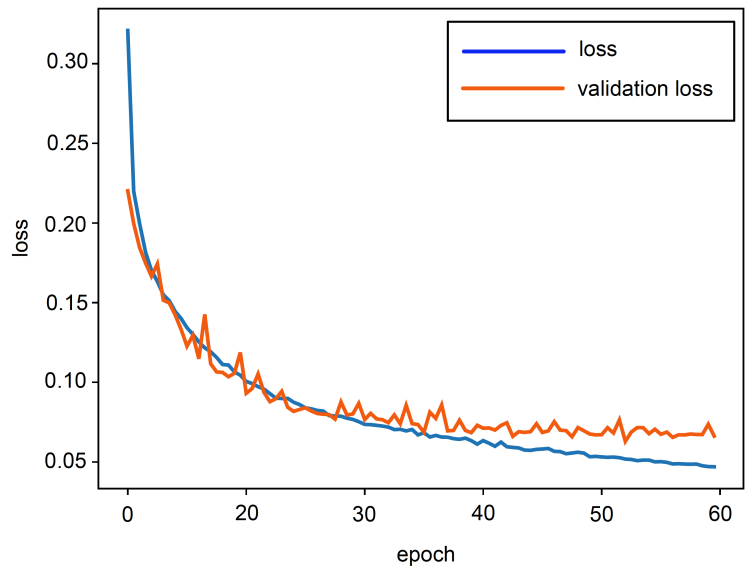

**Figure 3.** Learning curves for the model constructed in this study. The blue and red lines represent the training and validation losses, respectively.

**Table 1.** Classification Performance of Models with and without Data Augmentation

| Model | BACC | Event Type | Precision | Recall | F1 score |
|---|---|---|---|---|---|
| This study | 0.947 | A | 0.976 | 0.889 | 0.930 |
| | | B | 0.900 | 0.964 | 0.931 |
| | | Noise | 0.972 | 0.989 | 0.980 |
| CNN[a] | 0.921 | A | 0.949 | 0.855 | 0.899 |
| | | B | 0.869 | 0.935 | 0.901 |
| | | Noise | 0.952 | 0.975 | 0.963 |
| This study (Data Augumentation) | 0.968 | A | 0.984 | 0.935 | 0.959 |
| | | B | 0.940 | 0.975 | 0.957 |
| | | Noise | 0.982 | 0.993 | 0.987 |
| CNN[a] (Data Augumentation) | 0.910 | A | 0.977 | 0.784 | 0.870 |
| | | B | 0.826 | 0.962 | 0.889 |
| | | Noise | 0.953 | 0.984 | 0.968 |

[a]The CNN model was created using the method described in Nakano *et al*. (2022).

The evaluation metrics calculated were the Balanced Accuracy (BACC), Precision, Recall, and F1-score, as expressed as follows:

$$BACC: \frac{\frac{TP}{TP+FN}+\frac{TN}{TN+FP}}{2} \qquad (3)$$

$$Precision: \frac{TP}{TP+FP} \qquad (4)$$

$$Recall: \frac{TP}{TP+FN} \qquad (5)$$

$$F1\ Score: \frac{2TP}{2TP+FP+FN} \qquad (6)$$

where $TP$, $TN$, $FP$, and $FN$ represent true positives, true negatives, false positives, and false negatives, respectively. Compared with CNN-based models, our classification model demonstrated consistently higher classification performance across all metrics, underscoring its effectiveness in classifying volcanic earthquakes (Table 1). The recognition accuracy for the A-type was lower than that for the B-type for both methods, likely due to the smaller amount of training data available for the A-type. To address the imbalanced data issue, we applied data augmentation techniques. Specifically, we augmented A- and B-type earthquake data to match the number of noise samples (111,189) using two methods: (1) randomly adjusting the P-wave arrival times with random temporal shifts and (2) rotating the waveforms randomly by ±15º in three-dimensional space (involving EW, NS, and UD components). These methods preserved the waveform characteristics while introducing variability, enhancing the model's robustness (Supplemental Figure S2 illustrates an example of a rotated waveform). Data augmentation was applied only to the training set, and the validation set remained unaltered to accurately evaluate the model's generalization performance. The augmented dataset significantly improved classification performance compared to the imbalanced dataset, as listed in Table 1. For instance, the balanced accuracy (BACC) improved from 0.949 to 0.968, and the F1 scores for A-type, B-type, and noise also showed marked increases. In contrast, the CNN model exhibited a slight decrease in certain metrics, such as precision for the A-type, following data augmentation. This reduction may indicate underfitting of the CNN, as data augmentation increased the training set size without increasing the number of training epochs.

*AW Visualization*

Similar to the creation of many volcanic earthquake catalogs, the classification of earthquake types at Mount Asama was performed visually by humans based on the clarity of the P- and S-wave onsets and hypocenter depth. However, the classification processes of deep learning models are generally considered a "black-box," making it difficult to identify which features the model uses for classification. We therefore attempted to visualize which parts of the waveform the model attends to during classification by utilizing the AWs calculated within the model.

We explored a method for directly visualizing the AWs by extracting the attention weights from the multi-head self-attention mechanism within the Transformer encoder. Each Transformer layer consists of 10 attention heads; each head calculates its own set of AWs for every time step in the input sequence. The AW dimension ($297 \times 297$) corresponds directly to the sequence length entering the Transformer encoder, which is reduced from the original 3000 waveform samples down to 296 convolutional feature vectors, plus one CLS token, resulting in a total of 297 tokens. To obtain the final visualization, the AWs from all 10 heads are concatenated and then processed to compute the maximum value at each time step. This approach ensures that the most significant attention values are highlighted, providing a clear representation of where the model is focusing during the classification process.

The visualization results showed that AWs tended to concentrate mainly on the earthquake signal (Figure 4a). This suggests that our model focuses on features of seismic waveforms, such as how humans classify earthquakes. However, many cases were observed wherein the AW was dispersed across the entire waveform despite high confidence in the classification (Figure 4b).

We extracted the AW for two specific time intervals: 0.1 s before to 5 s after the P-wave, and from the start of the waveform to 1 s before the P-wave. The average AW within each interval was calculated, and the ratio of these averages was defined as the Attention Weight Ratio (AWR). This ratio provides insight into the relative concentration of AWs around the P-wave region compared to earlier parts of the waveform. The visualization showed that many values were concentrated at approximately 1, indicating the dispersion of the AW (Figure 4c).

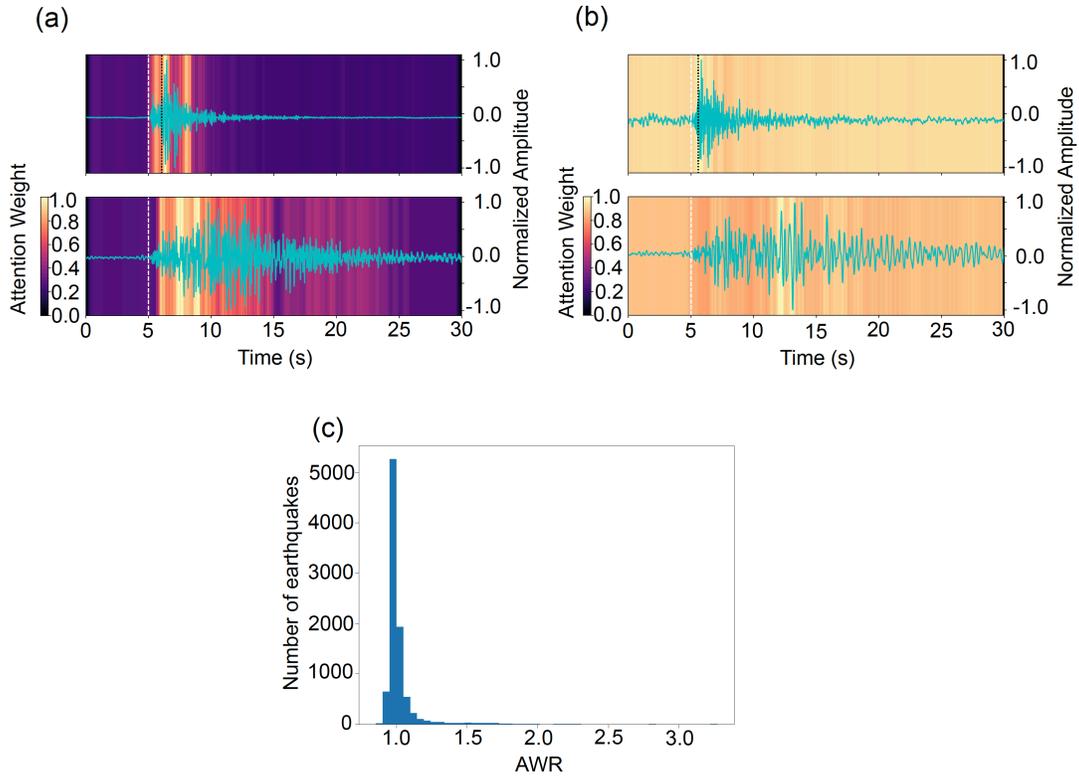

**Figure 4.** Prediction results from the validation data. (a) An example where a high AW is focused on the seismic signal. The upper figure shows a waveform classified as an A-type earthquake, and the lower figure shows one classified as a B-type earthquake. The AW is normalized between 0 and 1.0. (b) Examples of dispersed AW. The upper figure shows a waveform classified as an A-type earthquake, and the lower figure shows one classified as a B-type earthquake. The white- and black-dotted lines in the waveforms represent the onset of P- and S-waves selected by experts, respectively. (c) The number of earthquakes plotted against AWR.

When examining data with an AWR below 1.2, we identified several examples of ambiguous or potentially mislabeled events, as illustrated in Figure 5. Although we did not explicitly quantify the proportion across the entire dataset, these examples suggest that low-AWR data may contain waveforms that complicate clear classification. These examples represent typical cases of misclassifications observed in our analysis. They included cases of multiple earthquakes within a single dataset, time inversion errors, large non-earthquake signals, and earthquakes labeled as B-type that exhibit clear characteristics of A-type events, such as distinct P- and S-waves. Such data may lead to a dispersion of AWs across the waveform, diminishing the interpretability of the model by introducing noise. Additionally, we found that a significant subset of B-type earthquakes contained manually labeled S-wave arrivals that were not clear or typical for this type. This introduces further ambiguity in the classification process, suggesting the need for enhanced data quality control and potential automation in labeling to ensure the reliability of training data.

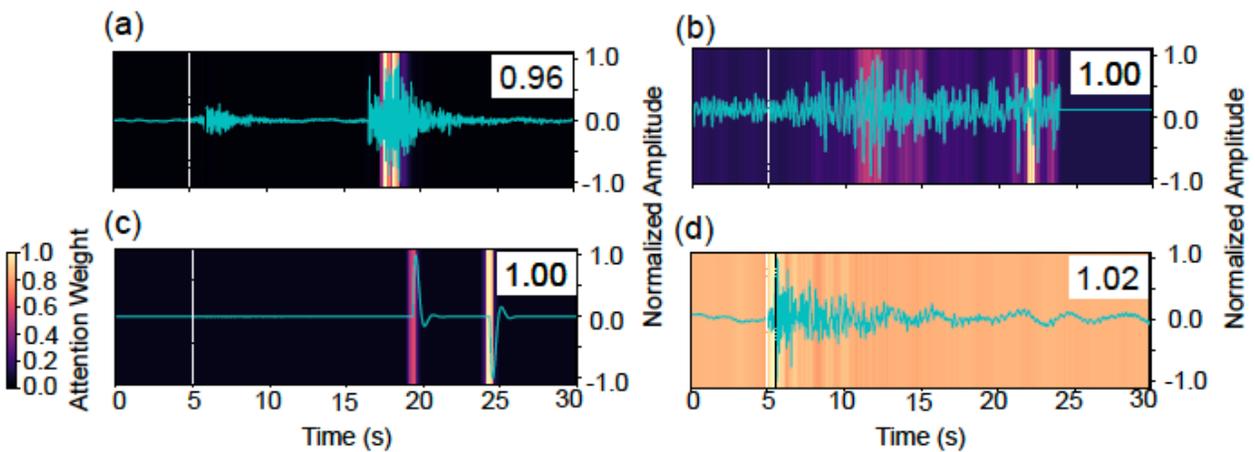

**Figure 5.** Examples of inappropriate data within the dataset, with the maximum AWR (Attention Weight Ratio) values displayed in the upper-right corner of each panel. These waveforms represent typical examples of misclassifications resulting from data-quality issues or ambiguous waveform characteristics. (a) A waveform containing multiple earthquakes within a single dataset. (b) A case of time-inverted data. (c) A waveform dominated by a large non-earthquake signal. (d) A waveform labeled as a B-type earthquake but exhibiting characteristics more consistent with an A-type event. Signal amplitude and AW are normalized between 0 and 1.0, respectively. The white- and black-dotted lines in the waveforms indicate the onset of P- and S-waves selected by experts, respectively.

**Discussion**

The quality of training data, which hinders effective learning, is presumed to be a significant reason for the divergence in AWs. Generally, B-type earthquakes are characterized by indistinct P- and S-waves. However, for Mount Asama, shallow earthquakes can be classified as B-type, regardless of their waveform characteristics. To investigate how data selection influences model performance, we conducted several experiments, including removing B-type events with S-wave labels, balancing the dataset through data augmentation, and training the model for four classes (A-type, B-type with S-wave, B-type without S-wave, and noise). These experiments were designed to address the challenges associated with data quality and imbalance, providing insights into the factors that affect classification accuracy and interpretability (Table 2). First, when B-type events with S-wave labels were removed, we observed an improvement in the AW distribution, which became more localized around the seismic signals (Figure 6a and b). Note that the waveforms presented in Figure 6a are identical to those shown previously in Figure 4b, illustrating how AW distributions changed due to the removal of B-type events labeled with S-wave arrivals. The increase in AWR could result from removing ambiguous labels; however, the decreased performance likely resulted from insufficient training due to fewer data points. Indeed, an additional experiment in which we randomly reduced waveforms to match the number before removing S-labeled events resulted in improved model performance, supporting the conclusion that there is a meaningful relationship between AWR and model accuracy. Supplemental Figure S3 shows separate histograms illustrating the distribution of Attention

Weight Ratio (AWR) values for A-type and B-type earthquakes. We found no clear or significant difference between the AWR distributions for these two event types, suggesting that the attention weight distributions are largely independent of earthquake type (A-type vs. B-type) in our dataset. The model trained on the modified dataset, even with data augmentation to balance the dataset, demonstrated lower performance compared to the original dataset that included S-wave labeled B-type events (Table 2). This suggests that removing S-wave labeled B-type events may inadvertently exclude important waveform characteristics, reducing the diversity and representativeness of the B-type class. Additionally, training the model with four classes resulted in poor classification accuracy for distinguishing between B-type events with and without S-waves. Detailed inspection revealed that some B-type events labeled with S-wave arrivals did not exhibit clearly distinguishable S-waves, potentially due to subjective labeling by human analysts (Figure 7). This overlap in waveform characteristics likely hindered the model's ability to effectively differentiate between the two subcategories of B-type events. These findings highlight the challenges of using subjective labels in training data. Note that while the validation datasets for A-type and Noise events were consistent across all experimental configurations, the validation data for B-type events differed in the four-class classification scenario, as B-type events were explicitly separated into those with and without S-wave labels (see Supplemental Table S1). Consequently, caution is required when directly comparing classification accuracy for the B-type class between three-class and four-class experiments.

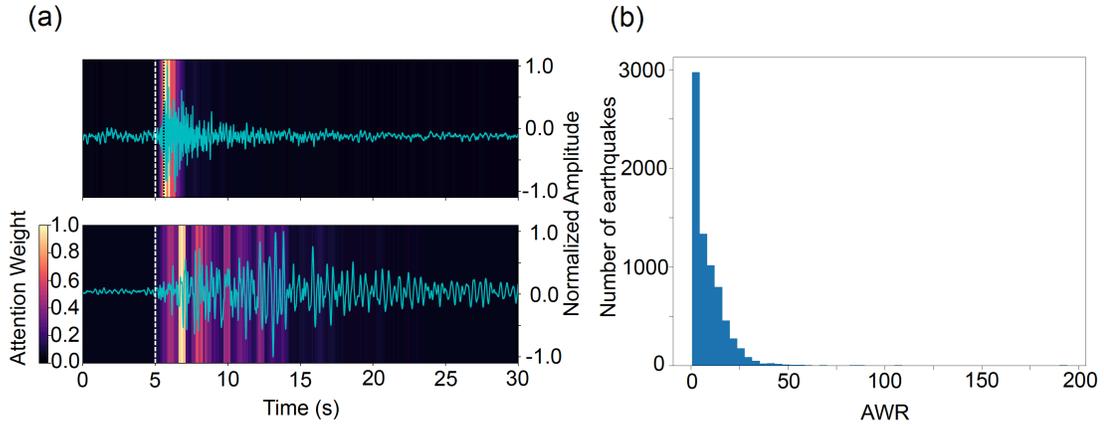

**Figure 6.** Evaluations using a model trained by excluding events with S-wave detection from B-type training data. (a) The same data are shown in Figure 4(b). (b) Number of earthquakes relative to the AWR.

**Table 2.** Classification Performance Across Different Data Configurations

| Model | Data without S [a] | | | Data without S with Data Augmentation [b] | | | 4 Class Classification [c] | | | |
|---|---|---|---|---|---|---|---|---|---|---|
| BACC | 0.938 | | | 0.957 | | | 0.780 | | | |
| Event Type | A | B | Noise | A | B | Noise | A | B with S | B without S | Noise |
| Precision | 0.904 | 0.950 | 0.961 | 0.942 | 0.972 | 0.958 | 0.934 | 0.749 | 0.569 | 0.956 |
| Recall | 0.934 | 0.888 | 0.992 | 0.951 | 0.929 | 0.990 | 0.865 | 0.434 | 0.832 | 0.987 |
| F1 score | 0.919 | 0.918 | 0.976 | 0.947 | 0.950 | 0.974 | 0.898 | 0.549 | 0.676 | 0.971 |

[a]Training data exclude B-type events with S-wave labels, with no data augmentation applied.
[b]Training data exclude B-type events with S-wave labels, with data augmentation applied to balance the dataset to the number of noise samples.

cTraining conducted using four categories: A-type, B-type with S-wave, B-type without S-wave, and noise.

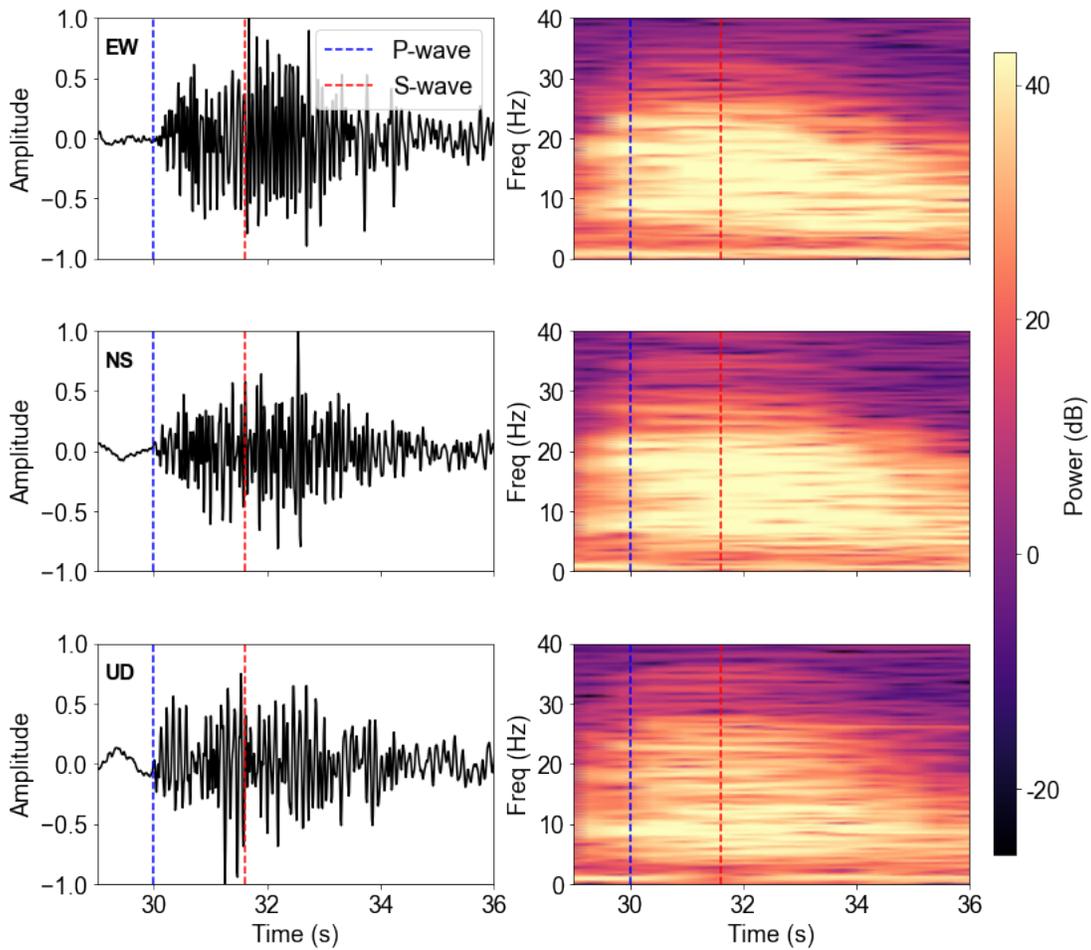

**Figure 7.** Example of a B-type event manually labeled with an S-wave arrival, which exhibits ambiguous or indistinct S-wave features. (a) Three-component (EW, NS, and UD) waveforms, each normalized to its maximum amplitude. Blue and red dashed lines represent manually labeled P- and S-wave arrivals, respectively. (b) Corresponding spectrograms (frequency range: 0–40 Hz) clearly illustrating spectral characteristics consistent with a typical B-type earthquake (cf. Figure 1c). Despite the manual S-wave label, both waveform and spectral content indicate indistinct S-wave energy, highlighting inherent ambiguities and subjectivity in traditional manual labeling approaches.

A slight tendency for a higher AWR at shorter source-receiver distances was observed (Figure 8a). As many earthquakes occur beneath the crater, stations within 3 km of the crater inevitably recorded more seismic events and may also have higher signal-to-noise ratios due to their proximity to shallow hypocenters. These factors collectively provided the Transformer architecture with richer waveform data, allowing it to better learn distinctive local features and thereby enhancing both classification performance and interpretability (Figure 8b). Furthermore, by overlaying the distribution of all data from each station with the distribution of data from sites with a prediction probability > 99%, these distributions matched. This trend remained unchanged even after removing the waveforms of events with large location errors which is defined in the previous section (Figure 8c).

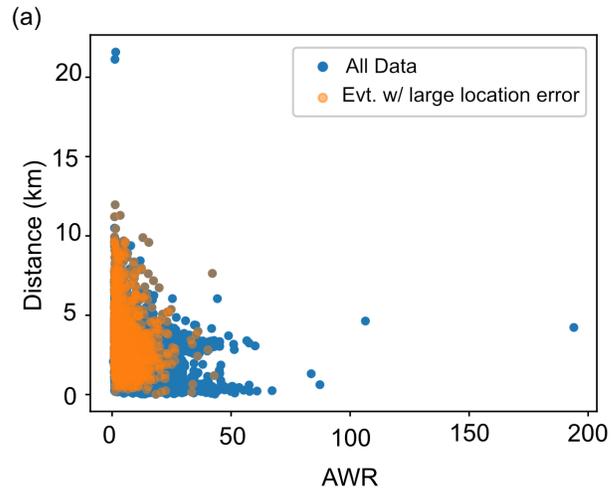

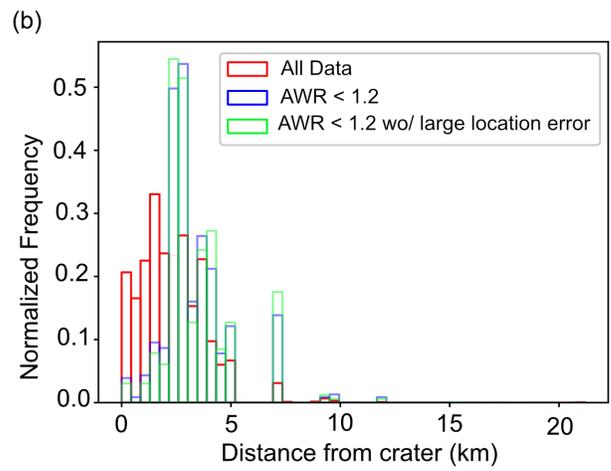

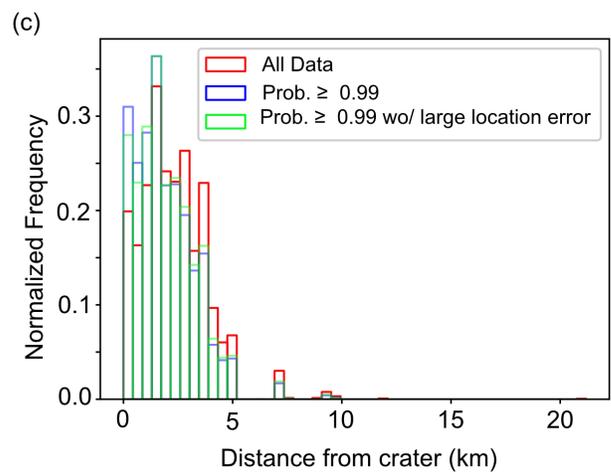

**Figure 8.** (a) Source-receiver distance relative to the AWR. Red circles indicate events with large location errors, while blue circles represent all other events. (b) Histograms of distances from the crater (Lat.: 36.4, Lon.: 138.52) to each station. Displays all test data (red), data with AWR < 1.2 (blue), and data with AWR < 1.2 excluding large location errors. Total data are normalized to sum to one. (c) Horizontal axis and red histogram are as described in (b). Blue represents the data histogram with a prediction probability > 99%, and green denotes the data histogram with a prediction probability > 99%, excluding events with large location errors. Total data are normalized to sum to one.

In this study, waveform classification was conducted independently at each seismic station. Although an explicit event-level classification integrating results from multiple stations was not performed, we confirmed that classifications were highly consistent across stations for individual events in our validation set. Developing a robust method for aggregating station-based classifications into event-level decisions will be addressed in future research.

Misclassifications are most likely to occur under certain conditions, such as high noise levels, ambiguous waveform characteristics, or overlapping features between classes. Such misclassifications could delay hazard assessments or lead to incorrect interpretations of volcanic processes. Therefore, it is essential to further analyze the conditions under which misclassifications arise. To this end, we analyzed confusion matrices generated from the testing data. The confusion matrix for the Transformer model trained with data augmentation (Table 3) shows that most misclassifications occurred between A- and B-type earthquakes. Specifically, 92 true B-type events were misclassified as A-type while 26 true A-type events were misclassified as B-type. These misclassifications likely stem from overlapping waveform characteristics between certain subcategories of A- and B-type earthquakes. Additionally, noise waveforms were classified with high accuracy, suggesting that the variability of noise characteristics had a limited impact on the model's performance in this dataset. For comparison, the confusion matrix for the CNN model (Table 3) reveals a higher rate of misclassification. Specifically, the CNN model misclassified 313 true A-type events as B-type and 47 true A-type as noise. This highlights the Transformer model's superior ability to handle the complexity of waveform data and its robustness against misclassification under similar conditions.

Table 3. Unified Confusion Matrices for Various Experimental Conditions*

| Model/Condition | Predicted A-type | Predicted B-type | Predicted Noise | Predicted B-type (with S) | Predicted B-type (without S) |
|---|---|---|---|---|---|
| **This Study (Data Augmentation)** | | | | | |
| True A-type | 1555 | 92 | 16 | | |
| True B-type | 26 | 1622 | 15 | | |
| True Noise | 0 | 12 | 1651 | | |
| **CNN (Data Augmentation)** | | | | | |
| True A-type | 1303 | 313 | 47 | | |
| True B-type | 29 | 1600 | 34 | | |
| True Noise | 2 | 25 | 1636 | | |
| **This Study (B-type without S)** | | | | | |
| True A-type | 1582 | 36 | 45 | | |
| True B-type | 91 | 1545 | 27 | | |
| True Noise | 7 | 9 | 1647 | | |
| **CNN (B-type without S)** | | | | | |
| True A-type | 1538 | 86 | 39 | | |
| True B-type | 115 | 1531 | 17 | | |
| True Noise | 11 | 18 | 1634 | | |
| **This Study (Four-class)** | | | | | |
| True A-type | 1438 | | 33 | 36 | 156 |
| True B-type (with S) | 61 | | 7 | 721 | 874 |
| True B-type (without S) | 38 | | 36 | 205 | 1384 |
| True Noise | 3 | | 1642 | 1 | 17 |
| **CNN (Four-class)** | | | | | |
| True A-type | 1371 | | 31 | 48 | 213 |
| True B-type (with S) | 59 | | 10 | 645 | 949 |
| True B-type (without S) | 48 | | 28 | 200 | 1387 |
| True Noise | 10 | | 1613 | 1 | 39 |

*Four-class models (Transformer and CNN) include the following categories: A-type, B-type with S-wave, B-type without S-wave, and Noise.

Notably, when the B-type events without S labelling were used, our model exhibited a slight increase in misclassification between A- and B-type events, compared with the original dataset (Table 3). This suggests that the exclusion of S-wave labeled B-type events may have reduced the diversity of the B-type class, making it more difficult for the model to generalize effectively. The CNN model's confusion matrix for the same dataset further emphasizes its difficulty in handling the reduced dataset, with a noticeable decline in classification performance (Table 3). This discrepancy may be attributed to differences in how the two architectures handle increased data variability introduced through augmentation. While the CNN model may have been more sensitive to the augmented data, potentially interpreting the variability as noise, Transformers benefited from the increased data diversity and volume, demonstrating their scalability with larger datasets (Kaplan et al., 2020). This property, combined with their ability to
34

effectively leverage augmented data, makes Transformers particularly well-suited for volcanic earthquake classification tasks. Addressing data imbalance through augmentation and employing architectures capable of harnessing the enhanced datasets are critical for improving classification accuracy in scenarios where available data for each class is highly imbalanced, such as in volcanic earthquake classification.

In the four-class classification experiment, misclassifications were concentrated between S-wave labeled and unlabeled B-type events (Table 3). This result aligns with earlier discussion and highlights the need for further waveform scrutiny and feature validation to effectively distinguish between these subcategories. These results collectively emphasize the importance of balancing data quality and diversity in building robust classification models. While the inclusion of S-wave labeled B-type events introduces noise, it also contributes essential variability that aids in generalization. Addressing the overlap between subcategories through more objective feature extraction methods, such as autoencoders, could reduce subjectivity and improve classification accuracy. Integrating clustering methods or connecting the bottleneck layer of an autoencoder to a Transformer encoder could further enhance the interpretability and performance of classification models (e.g., Kong *et al.*, 2021). In future research, employing autoencoder-based methods may allow for effective extraction of essential features even from sparse datasets, facilitating the development of volcano-specific classification models. This approach could further reduce the human and time resources required for manual labeling, thus significantly enhancing the efficiency and reproducibility of volcanic earthquake classification workflows. In addition, clarifying these subtle distinctions explicitly not only highlights the subjective challenges inherent in manual labeling but also underscores the broader necessity of developing objective, reproducible classification methods. This clarity can directly benefit future studies and practical earthquake monitoring efforts by promoting consistent, reliable classification and deeper mechanistic insights.

Finally, our study demonstrated the effectiveness of using a Transformer encoder for volcanic earthquake classification. However, applying this model to other volcanic regions presents additional challenges. As the model incorporates features influenced by geological structures, source-receiver distances, and source characteristics, it may require retraining or adaptation through transfer learning to ensure effectiveness in new regions. Furthermore, achieving comparable performance in new regions will likely necessitate large datasets or data augmentation using synthetic or cross-regional data.

To evaluate the suitability of spectral metrics such as Frequency Index (FI) for distinguishing volcanic earthquake types, we examined the FI distributions of manually labeled A-type and B-type earthquakes (Figure 9). Following Yukutake et al. (2023), which builds upon the definitions proposed by Buurman et al. (2010) and Matoza et al. (2014), we calculated FI as the logarithmic ratio of mean spectral amplitudes between high-frequency (10–15 Hz) and low-frequency (1–4 Hz) bands, derived from the Fourier transforms of vertical-component waveforms. Although this definition is commonly used, FI performance in our dataset might potentially improve by evaluating a higher-frequency band, such as 20–30 Hz or 30–40 Hz, as energy is notably depleted above 30 Hz (Figure 2). Nevertheless, significant overlap observed in FI values clearly indicates that using a spectral metric alone is insufficient for robust earthquake classification at Mount Asama. As explicitly illustrated in Figure 9, some B-type events exhibit frequency indices overlapping with A-type events, and the FI distribution for B-type events appears bimodal, concentrating around FI values of approximately -1.3 and -0.2. This bimodal distribution may reflect inherent variability within B-type earthquakes themselves, possibly due to differences in source processes or propagation paths. Additionally, classification practices at Mount Asama contribute significantly to this overlap. Human analysts at Mount Asama classify shallow earthquakes beneath the crater primarily based on hypocenter depth and event location, rather than strictly waveform clarity or frequency characteristics. Consequently, even events exhibiting clearer P- and S-wave arrivals and higher FI values—features typically associated with A-type events—are sometimes classified as B-type. This practice introduces inherent ambiguity and complexity into classification, further contributing to the observed overlap between these two event types. Thus, waveform characteristics among events with similar FI values can differ substantially, highlighting the difficulty of using frequency-domain information alone for robust classification.

Our Transformer-based model addresses this limitation by leveraging waveform characteristics beyond simple spectral ratios. Specifically, the Transformer extracts and utilizes nonlinear and temporal waveform features, such as envelope shape, duration, and arrival patterns of seismic phases, which are critical for accurate classification but are not represented adequately by FI or similar single-parameter spectral indices. Consequently, our approach achieves high classification performance, as evidenced by high F1 scores of 0.959 and 0.957 for A-type and B-type events, respectively, clearly demonstrating the benefit of employing deep learning techniques to classify volcanic earthquakes.

Finally, our results highlight the broader relevance of advanced AI methodologies, such as Transformer architectures, in volcanic seismology. Deep learning models with interpretability tools (e.g., attention visualizations) can systematically reveal ambiguities and inconsistencies in traditional earthquake classifications, thus guiding improvements in data quality and labeling practices (Mousavi et al., 2020; Kong et al., 2021). Nevertheless, operational implementation of these AI techniques requires careful validation and clear demonstration of interpretability to ensure trust and practical utility in volcanic monitoring contexts (Rudin, 2019). Future studies should further refine these approaches and rigorously evaluate their applicability to diverse volcanic environments.

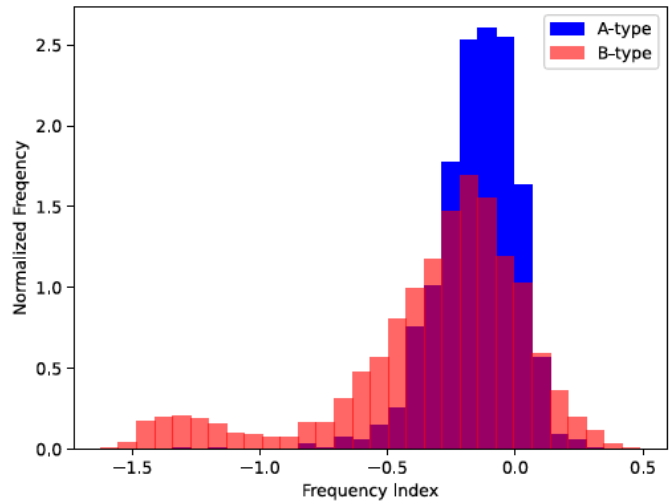

**Figure 9.** Frequency Index (FI) distributions for manually labeled A-type (blue) and B-type (red) earthquakes at Mount Asama, shown as probability density functions (PDFs). Histograms are normalized so that the total area under each equals 1, allowing clear visual comparison of distribution shapes. FI was calculated following Yukutake et al. (2023), which builds upon the definitions proposed by Buurman et al. (2010) and Matoza et al. (2014), using the logarithmic ratio of mean spectral amplitudes between high-frequency (10–15 Hz) and low-frequency (1–4 Hz) bands, as described explicitly in the Discussion. The significant overlap indicates the difficulty of accurately classifying earthquake types based solely on FI.

**Conclusion**

In this study, we developed a Transformer-based deep learning model capable of accurately classifying volcanic earthquakes at Mount Asama into A-type, B-type, and noise categories, achieving high F1 scores (~0.96) that significantly surpass conventional CNN-based methods. Our model effectively handled imbalanced data through augmentation strategies, substantially enhancing classification accuracy.

Visualization of Attention Weights (AW) revealed that the Transformer model consistently attends to physically meaningful waveform features, primarily around P-wave arrivals, thereby providing interpretability and insight into the model's decision-making process. Analysis of the Attention Weight Ratio (AWR) highlighted critical issues with data quality, such as ambiguous labeling and waveform inconsistencies, serving as a practical tool to identify problematic data.

Our results also underscore the importance of strategically positioned seismic stations near volcanic craters, significantly enhancing model accuracy and interpretability. Ensuring high-quality, consistent datasets is essential for reliable automated volcanic earthquake classification. By systematically addressing ambiguity in manual classification, our Transformer-based approach provides objective and reproducible criteria, benefiting volcano monitoring operations by reducing subjectivity and enhancing efficiency. Future research, including advanced waveform feature extraction methods and robust transfer learning frameworks, is needed to expand this approach's applicability to other volcanic regions.

**Data and Resources**

Waveform data were obtained from Hinet (https://hinetwww11.bosai.go.jp/auth/?LANG=en), the Earthquake Research Institute, and the Japan Meteorological Agency via the JVDN system (https://jvdn.bosai.go.jp/app/pages/index.html?root=publicDataList&lang=en). The model developed in this study, training data, and seismic station lists were downloaded from our GitHub page (https://github.com/yugosuz/Volcanic-Earthquake-Classification-using-Transformer-Encoder).

**Declaration of Competing Interests**

The authors declare no competing interests.

**Acknowledgments**

We sincerely thank the Editor-in-Chief, Dr. Allison Bent, and three anonymous reviewers for their constructive and valuable feedback, which significantly improved our manuscript. The waveform data and earthquake catalogs used in this study were collected and created by the Asama Volcano Observatory at the University of Tokyo. Waveforms were recorded by stations operated by the ERI, NIED, and JMA. Maps were prepared using Geophysical Mapping Tools (Wessel et al., 2019), and other figures were created using Matplotlib, a Python module. This study was supported by JSPS KAKENHI (Grant Nos. 22K03752, 15KK0171) and the Next Generation Volcano Research and Human Resource Development Project of the Ministry of Education, Culture, Sports, Science, and Technology of Japan.

**References**

Bebbington, M. S. (2007). Identifying volcanic regimes using Hidden Markov Models. Geophysical Journal International, 171(2), 921–942. https://doi.org/10.1111/j.1365-246X.2007.03545.x

Buurman, H., M. E. West, J. A. Power, and M. L. Coombs (2006). Seismic precursors to volcanic explosions during the 2006 eruption of Augustine Volcano, *U.S. Geological Survey Professional Paper 1769*, 41–57.

Canário, J. P., R. Mello, M. Curilem, F. Huenupan, and R. Rios (2020). In-depth comparison of deep artificial neural network architectures on seismic events classification, *J. Volcanol. Geotherm. 401*, 106881. doi:10.1016/j.jvolgeores.2020.106881.

Chouet, B. (1988). Resonance of a fluid-driven crack: radiation properties and implications for the source of long-period events and harmonic tremor, *J. Geophys. Res.: Solid Earth. 93*, 4375–4400. doi: 10.1029/JB093iB05p04375.

Chouet, B. A. (1996). Long-period seismicity: its source and use in eruption forecasting, Nature 380, 309–316.

Chouet, B. A. (2003). Volcano seismology, Pure Appl. Geophys. 160, 739–788.

Gulati, A., J. Qin, C. Chiu, N. Parmar, Y. Zhang, J. Yu, W. Han, S. Wang, Z. Zhang, Y. Wu, and R. Pang (2020). Conformer: Convolution-augmented transformer for speech recognition, doi: 10.48550/arXiv.2005.08100.

Huang, Y. (2018). Earthquake rupture in fault zones with along-strike material heterogeneity. Journal of Geophysical Research: Solid Earth, 123(11), 9884–9898. https://doi.org/10.1029/2018JB016354

Ida, Y., and Ishida, M. (2022). Analysis of seismic activity using self-organizing map: Implications for earthquake prediction. Geophysical Journal International, 229(2), 1077–1091. https://doi.org/10.1093/gji/ggab476

Iguchi, M. (2013). Magma movement from the deep to shallow Sakurajima Volcano as revealed by geophysical observations, *Bull. Volc. Soc. Japan 58*(1), 1–18.

Kaplan, J., S. McCandlish, T. Henighan, T. B. Brown, B. Chess, R. Child, S. Gray, A. Radford, Kingma, D., and J. L. Ba (2014). Adam: A method for stochastic optimization. *Comp Sci*, 1-15. https://arxiv.org/abs/1412.6980

Köhler, C. M., Kouwenhoven, T. J., and Kuijken, K. (2010). The effect of the Galactic tide on local stellar streams. Astronomy & Astrophysics, 519, A13. https://doi.org/10.1051/0004-6361/201014147

Kong, Q., A. Chiang, A. C. Aguiar, M. G. Fernández-Godino, S. C. Myers, and D. D. Lucas (2021). Deep convolutional autoencoders as generic feature extractors in seismological applications, *Artif. Intell. Geosci., 2,* 96-106, doi: 10.1016/j.aiig.2021.12.002.

Langer, H., Falsaperla, S., and Campanini, R. (2011). Automatic classification and identification of volcanic tremor sources based on support vector machines. Geophysical Research Letters, 38(19), L19304. https://doi.org/10.1029/2011GL049065

Langer, H., Falsaperla, S., Masotti, M., Campanini, R., Spampinato, S., and Messina, A. (2009). Synopsis of supervised and unsupervised pattern classification techniques applied to volcanic tremor data at Mt Etna, Italy. Geophysical Journal International, 178(2), 1132–1144. https://doi.org/10.1111/j.1365-246X.2009.04211.x

Lara, F., R. Lara-Cueva, J. C. Larco, E. V. Carrera, and R. León (2021). A deep learning approach for automatic recognition of seismo-volcanic events at the Cotopaxi volcano, *J. Volcanol. Geotherm. 409*, 107142. doi:10.1016/j.jvolgeores.2020.107142.

Li, R., Z. Mai, Z. Zhang, J. Jang, and S. Sanner (2023). TransCAM: Transformer attention-based CAM refinement for weakly supervised semantic segmentation, *J. Vis. Commun. Image Represent. 92*, 103800, doi: 10.1016/j.jvcir.2023.103800.

Loshchilov, I. (2017). Decoupled weight decay regularization, aXiv preprint arXiv:1711.05101.2017

Malfante, M., M. D. Mura, J. I. Mars, J. P. Metaxian, O. Macedo, and A. Inza (2018). Automatic classification of volcano seismic signatures, *J. Geophys. Res.: Solid Earth 123*, 10,645–10,658.

Matoza, R. S., P. M. Shearer, and P. G. Okubo (2014). High‐precision relocation of long‐period events beneath the summit region of Kīlauea Volcano, Hawaiʻi, from 1986 to 2009, Geophys. Res. Lett. 41 (10), 3413–3421.

McNutt, S. R. (1996). Seismic monitoring and eruption forecasting of volcanoes: A review of the state-of-the-art and case histories, *Monitoring and Mitigation of Volcano Hazards* 99–146, Berlin, Heidelberg: Springer. doi: 10.1007/978-3-642-80087-0_3

Minakami, T., S. Utibori, S. Hiraga, T. Miyazaki, N. Gyoda, and T. Utsunomiya (1970). Seismometrical studies of Volcano Asama, Part I. Seismic and volcanic activities of Asama during 1934-1969, *Bull. Earthq. Res. Inst. 48*, 235–301.

Mousavi, S. M., W. L. Ellsworth, W. Zhu, L. Y. Chuang, and G. C. Beroza (2020). Earthquake transformer—an attentive deep-learning model for simultaneous earthquake detection and phase picking, Nat. Commun. 11, 3952.

Nakano, M., and D. Sugiyama (2022). Discriminating seismic events using 1D and 2D CNNs: applications to volcanic and tectonic datasets, *Earth Planets Space*, *74*, 134. doi: 10.1186/s40623-022-01696-1.

Nishimura, T., and M. Iguchi (2011). Volcanic earthquakes and tremor in Japan, *Kyoto University Press*. ISBN: 4876989796, 9784876989799.

Oikawa, J., Y. Ida, and H. Tsuji (2006). Precise hypocenter determination for volcanic earthquakes observed at Asama Volcano, Japan, *Bull. Volc. Soc. Japan* 51(2), 117–124.

Permana, T., Nishimura, T., Nakahara, H., & Shapiro, N. (2022). Classification of volcanic tremors and earthquakes based on seismic correlation: Application at Sakurajima volcano, Japan. Geophysical Journal International, 229(2), 1077–1091. https://doi.org/10.1093/gji/ggab476

Power, J. A., and D. C. Roman (2024). Event classification, seismicity, and eruption forecasting at Great Sitkin Volcano, Alaska: 1999–2023, *J. Volcanol. Geotherm. Res.* **454**, 108182.

Ribeiro, M. T., S. Singh, C. Guestrin (2016), "Why should I trust you?": Explaining the predictions of any classifier, KDD '16: Proceedings of the 22nd ACM SIGKDD International Conference on Knowledge Discovery and Data Mining, 1135-1144, doi: 10.1145/2939672.2939778

Rudin, C. (2019). Stop explaining black box machine learning models for high stakes decisions and use interpretable models instead, Nat. Mach. Intell. 1, 206–215.

Takeo, M., Y. Aoki, and T. Koyama (2022). Recent volcanic activity at the Asama volcano and long-period seismic signals, *Proc. Jpn. Acad. Series B*, *98*(8), 416–438. doi: 10.2183/pjab.98.022.

Titos, M., A. Bueno, L. Garcia, and C. Benitez (2018). A deep neural networks approach to automatic recognition systems for volcano-seismic events, *IEEE J. Sel. Top. Appl. Earth Obs. Remote Sens. 11*(5), 1533–1544, doi: 10.1109/JSTARS.2018.2803198

Vaswani, A., N. Shazeer, N. Parmar, J. Uszkoreit, L. Jones, A.N. Gomez, L. Kaiser, and I. Polosukhin (2017). Attention is all you need, *Adv. Neural. Inf. Process. Syst. 30*, 6000–6010.

Wessel, P., J. F. Luis, L. Uieda, R. Scharroo, F. Wobbe, W. H. F. Smith, and D. Tian (2019). The generic mapping tools version 6, *Geochem. Geophys. 20*(11), 5556–5564, doi: 10.1029/2019GC008515.

Wu, J., and D. Amodei (2020). Scaling laws for neural language models, doi: 10.48550/arXiv.2001.08361.

Yukutake, Y., A. Kim, and T. Ohminato (2023). Reappraisal of volcanic seismicity at the Kirishima volcano using machine learning, *Earth Planets Space* **75**, 183.

**Full mailing address for each author**

Yugo Suzuki

Yokohama City University, 22 Seto Kanazawa-ku Yokohama city Kanagawa prefecture Japan, 236-0027

Yohei Yukutake

Earthquake Research Institute, The University of Tokyo, 1-1-1 Yayoi Bunkyo-ku, Tokyo, Japan, 113-0032

Takao Ohminato

Earthquake Research Institute, The University of Tokyo, 1-1-1 Yayoi Bunkyo-ku, Tokyo, Japan, 113-0032

Masami Yamasaki

Yokohama City University, 22 Seto Kanazawa-ku Yokohama city Kanagawa prefecture Japan, 236-0027

Ahyi Kim

Yokohama City University, 22 Seto Kanazawa-ku Yokohama city Kanagawa prefecture Japan, 236-0027

**Figure Captions**

**Figure 1.** Network architecture. Three-component seismic waveforms are passed through a one-dimensional convolution layer (kernel size: 50; input channel size: 3; output channel size: 150; stride: 10), followed by batch normalization, ReLU activation, and positional embedding. A classification token (CLS) is added to the input sequence and processed through three stacked Transformer encoder layers, each consisting of 10 multi-head attention mechanisms, feedforward sublayers, and residual connections. The CLS token output is extracted and passed through a linear transformation to classify A- and B-type events, as well as noise.

**Figure 2.** Distribution of earthquakes and stations and examples of input data. Earthquakes detected between January 2003 and October 2022, only with location determination errors less than 100 m in horizontal and depth directions (based on the earthquake catalog from the Earthquake Research Institute, The University of Tokyo), were plotted. (a) Distribution of earthquakes (circles) and stations (blue triangles) at Mount Asama. Red circles represent A-type events, green circles represent B-type labeled events, and open circles represent earthquakes that do not belong to either type. The top image is a map view, and the bottom image is a cross-sectional view along the red line in the map view. In the cross-section, "Height" represents elevation relative to sea level, with positive values indicating depth below the surface. The "0" on the vertical axis indicates the surface level. (b) An example of an A-type earthquake. The top panel shows the seismic waveform, and the bottom image shows the spectrogram. (c) Same as (b), but representing an example of a B-type earthquake.

**Figure 3.** Learning curves for the model constructed in this study. The blue and red lines represent the training and validation losses, respectively.

**Figure 4.** Prediction results from the validation data. (a) An example where a high AW is focused on the seismic signal. The upper figure shows a waveform classified as an A-type earthquake, and the lower figure shows one classified as a B-type earthquake. The AW is normalized between 0 and 1.0. (b) Examples of dispersed AW. The upper figure shows a waveform classified as an A-type earthquake, and the lower figure shows one classified as a B-type earthquake. The white- and black-dotted lines in the waveforms represent the onset of P- and S-waves selected by experts, respectively. (c) The number of earthquakes plotted against AWR.

**Figure 5.** Examples of inappropriate data within the dataset, with the maximum AWR (Attention Weight Ratio) values displayed in the upper-right corner of each panel. These waveforms represent typical examples of misclassifications resulting from data-quality issues or ambiguous waveform characteristics. (a) A waveform containing multiple earthquakes within a single dataset. (b) A case of time-inverted data. (c) A waveform dominated by a large non-earthquake signal. (d) A waveform labeled as a B-type earthquake but exhibiting characteristics more consistent with an A-type event. Signal amplitude and AW are normalized between 0 and 1.0, respectively. The white- and black-dotted lines in the waveforms indicate the onset of P- and S-waves selected by experts, respectively.

**Figure 6.** Evaluations using a model trained by excluding events with S-wave detection from B-type training data. (a) The same data are shown in Figure 4(b). (b) Number of earthquakes relative to the AWR.

**Figure 7.** Example of a B-type event manually labeled with an S-wave arrival, which exhibits ambiguous or indistinct S-wave features. (a) Three-component (EW, NS, and UD) waveforms, each normalized to its maximum amplitude. Blue and red dashed lines represent manually labeled P- and S-wave arrivals, respectively. (b) Corresponding spectrograms (frequency range: 0–40 Hz) clearly illustrating spectral characteristics consistent with a typical B-type earthquake (cf. Figure 1c). Despite the manual S-wave label, both waveform and spectral content indicate indistinct S-wave energy, highlighting inherent ambiguities and subjectivity in traditional manual labeling approaches.

**Figure 8.** (a) Source-receiver distance relative to the AWR. Red circles indicate events with large location errors, while blue circles represent all other events. (b) Histograms of distances from the crater (Lat.: 36.4, Lon.: 138.52) to each station. Displays all test data (red), data with AWR < 1.2 (blue), and data with AWR < 1.2 excluding large location errors. Total data are normalized to sum to one. (c) Horizontal axis and red histogram are as described in (b). Blue represents the data histogram with a prediction probability > 99%, and green denotes the data histogram with a prediction probability > 99%, excluding events with large location errors. Total data are normalized to sum to one.

**Figure 9.** Frequency Index (FI) distributions for manually labeled A-type (blue) and B-type (red) earthquakes at Mount Asama, shown as probability density functions (PDFs). Histograms are normalized so that the total area under each equals 1, allowing clear visual comparison of distribution shapes. FI was calculated following Yukutake et al. (2023), which builds upon the definitions proposed by Buurman et al. (2010) and Matoza et al. (2014), using the logarithmic ratio of mean spectral amplitudes between high-frequency (10–15 Hz) and low-frequency (1–4 Hz) bands, as described explicitly in the Discussion. The significant overlap indicates the difficulty of accurately classifying earthquake types based solely on FI.